\documentclass{article}
\usepackage{ijcai23}

\usepackage{times}
\usepackage{soul}
\usepackage{url}
\usepackage[hidelinks]{hyperref}
\usepackage[utf8]{inputenc}
\usepackage[small]{caption}
\usepackage{graphicx}
\usepackage{amsmath}
\usepackage{amsthm}
\usepackage{booktabs}
\usepackage{algorithm}
\usepackage{algorithmic}
\usepackage{comment}
\usepackage[switch]{lineno}
\usepackage{tabularray}

\usepackage [autostyle, english = american]{csquotes}
\MakeOuterQuote{"}

\title{Evaluating Human-AI Interaction via Usability, User Experience and Acceptance Measures for MMM-C: A Creative AI System for Music Composition}

\author{
Renaud Bougueng Tchemeube$^1$
\and
Jeffrey Ens$^1$\and
Cale Plut$^1$\and
Philippe Pasquier$^1$\and \\
Maryam Safi$^2$\and
Yvan Grabit$^2$\And
Jean-Baptiste Rolland$^2$
\affiliations
$^1$Simon Fraser University, \\
$^2$Steinberg Media Technologies GmbH\\
\emails
\{rbouguen, cplut, pasquier\}@sfu.ca,
jeffreyjohnens@gmail.com,
\{m.safi, y.grabit, jb.rolland\}@steinberg.de
}

\begin{document}

\maketitle

\begin{abstract}
    With the rise of artificial intelligence (AI), there has been increasing interest in human-AI co-creation in a variety of artistic domains including music as AI-driven systems are frequently able to generate human-competitive artifacts. Now, the implications of such systems for musical practice are being investigated. We report on a thorough evaluation of the user adoption of the Multi-Track Music Machine (MMM) as a co-creative AI tool for music composers. To do this, we integrate MMM into Cubase, a popular Digital Audio Workstation (DAW) by Steinberg, by producing a "1-parameter" plugin interface named MMM-Cubase (MMM-C), which enables human-AI co-composition. We contribute a methodological assemblage as a 3-part mixed method study measuring usability, user experience and technology acceptance of the system across two groups of expert-level  composers: hobbyists and professionals. 
    Results show positive usability and acceptance scores. Users report experiences of novelty, surprise and ease of use from using the system, and limitations on controllability and predictability of the interface when generating music. 
    Findings indicate no significant difference between the two user groups.
\end{abstract}

\section{Introduction}
\begin{figure}
\includegraphics[width=\linewidth]{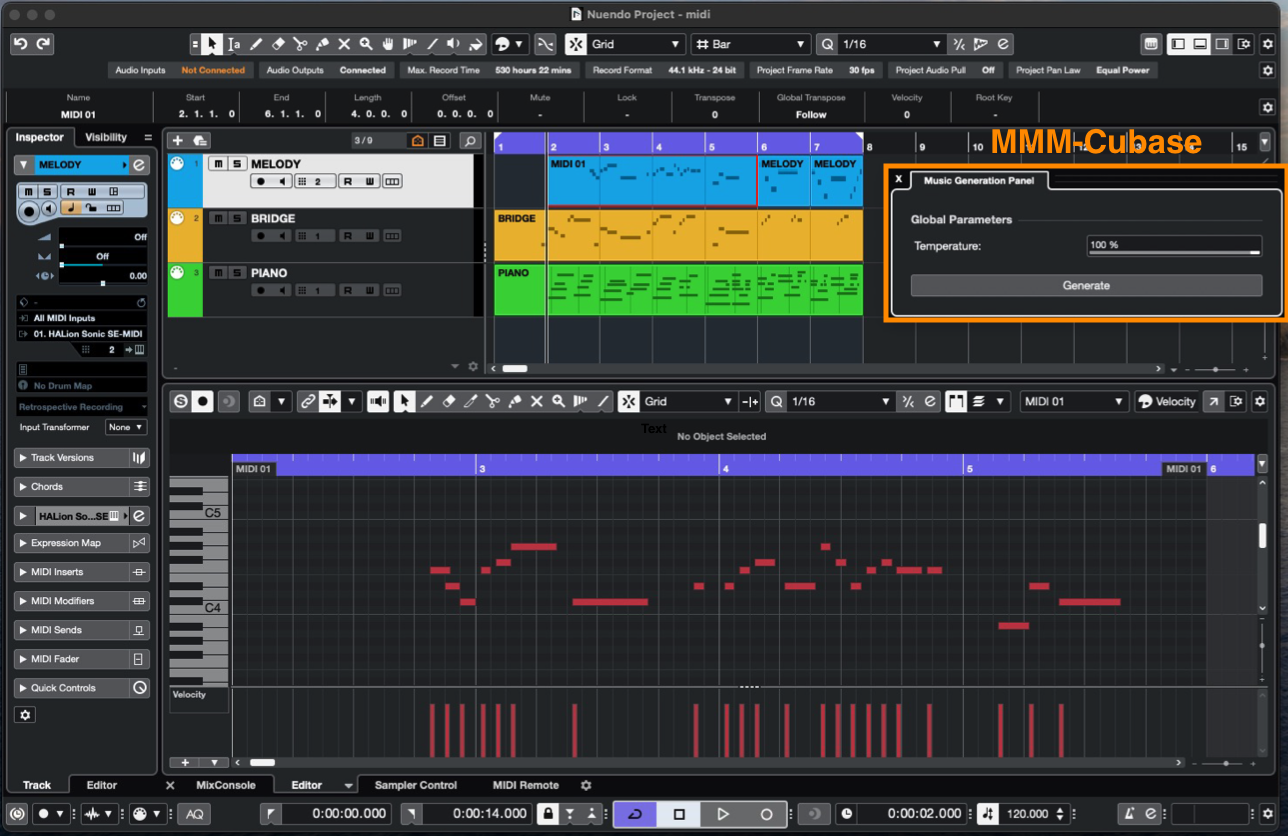}
\caption{MMM-Cubase's Interface in Cubase}
\label{fig:mmmincubase}
\end{figure}
Prompted by significant advancements in the field of Artificial Intelligence (AI), there has been renewed interest in human-AI co-creation in a variety of domains, including drawing \cite{davis2016empirically}, writing \cite{clark2018creative}, video game generation \cite{guzdial2019friend}, sound design \cite{thorogood2013computationally,kranabetter2022audio}, animation \cite{alemi2017walknet,alemi2017groovenet} and music composition \cite{louie2020novice}. 
One of the core challenges is to develop systems that can effectively facilitate user goals. Although a plethora of music composition systems have been developed \cite{donahue2019lakhnes,ens2020mmm,liang2017automatic,oore2020time,roberts2018hierarchical}, much of this research has focused on generative models as the primary end-goal, rather than improving the affordances in designing their interfaces for practical scenarios \cite{sturm2019machine}. To better address these challenges, Ens et al. develop the Multi-Track Music Machine (MMM) \cite{ens2020flexible}, a machine learning (ML) music system capable of generating multi-track symbolic music in a controlled manner. MMM is a powerful and highly controllable generative transformer model with the ability to fully instruct for melody, harmony and rhythmic generation of new musical patterns. The model uses a unique data representation where a MIDI file is encoded as a single sequence of concatenated MIDI events from each track at a time. It is trained on the MetaMIDI dataset \cite{ens2021building}, a collection of half a million MIDI files of various music genres.
In contrast to self-contained systems which generate an entire musical piece without human intervention \cite{musenet,huang2018music,donahue2019lakhnes,roberts2018hierarchical}, MMM is designed to be integrated into the composition workflow, enabling three types of user action: track in-filling, bar in-filling, and attribute controls such as instrument type, note density, polyphony, or note length among many others. 

Evaluating AI systems, both predictive and generative ones, has been a growing topic of interest within the AI research community, particularly because of the challenges posed by designing effective interfaces around, often opaque, AI behavior. As such systems are being rapidly brought up into the public domain and made accessible to various user groups, it becomes imperative to further our understanding of the impact on  human factors, of the novel interactive processes they afford. In the areas of computer-assisted composition (CAC) and music generation, many studies typically focus on novices with standalone "demo" systems, often built to showcase algorithmic capabilities of the models. 
%
In this paper, we present a \textit{mixed methods} research attempting to characterize the nature of human-AI co-creative interaction by evaluating \textit{usability}, \textit{user experience} and \textit{acceptance} of expert composers, in the practical context of \textit{MMM-Cubase} (MMM-C) (Figure \ref{fig:mmmincubase}), a "1-parameter" plugin interface of  MMM, the style-agnostic highly controllable multi-track generative model. 
Our proposed methodological assemblage is unique and more specific than previous work, 
and results from this minimal MMM-C's interface acts as a baseline for comparison and benchmarking, with more complex interfaces that have/will employ the MMM model (e.g., Calliope \cite{bougueng2022calliope}).
We seek to add to the body of literature with expert composers' insights and to understand the level of affordance necessary to make the model usable in a practical workflow. 
Our research questions are as follow:

\begin{itemize}
    \item RQ1: is MMM usable and effective at generating what is being requested while helping produce creative outputs for expert composers?

    \item RQ2: How do expert composers perceive their sense of autonomy, flexibility, accessibility, and authorship with regards to using MMM in Cubase?
    
    \item RQ3: What is the level of technology acceptance of expert composers regarding the use of MMM in Cubase?
\end{itemize}

We choose a commercial DAW for our methodology design to replicate the composition environment typical for expert composers. Cubase was selected over other options as it is one of the oldest and most popular ones, with a long history of expert use, and a dedicated user base to pool from.
The reader may contact the authors for instructions on how to navigate the beta portal at \url{https://beta.steinberg.net} for accessing the MMM-C research prototype.

\section{Background}
\label{background}
Musical Metacreation (MuMe) \cite{pasquier2016introduction} is a field of research which addresses the partial or complete automation of creative musical tasks including composition, interpretation, improvisation, accompaniment or mixing. It investigates purely generative systems for music as well as interactive ones. Computer-Assisted Composition (CAC) is a subfield that focuses on developing systems for automating music composition processes; namely exploration, development and rendering of musical ideas. There exists several compositional tasks a system can address: multi-track pattern generation, multi-track complete generation, rhythm generation, harmonization, chord progression generation, melody generation, interpolation, form-filling and orchestration. Each of these tasks can be realized given a conditioning or not on prior musical sequences, on supporting instrumentation, or generative control parameters.
Achieving generative capabilities often requires the usage of ML-based statistical modeling techniques \cite{briot2017deep};
a broad range of algorithmic techniques such as Markov models, rule-based, factor oracles, neural networks, deep learning and more. These algorithms are capable of generating new musical content. Software systems are built to interface them to users for interactivity, and in some cases, for co-creativity.
Co-creativity refers to "multiple parties contributing to the creative process in a blended manner" \cite{davis2013human}. Traditionally, HCI takes place in such a way that tasks accomplished by the user and the computer follow a distribution of labor \cite{kantosalo2015interaction}. The user completes a high-level task by identifying and requesting for low-level computable tasks to be completed by the system. 
Co-creativity goes beyond this paradigm and allow all parties to collaboratively contribute synthetic results e.g., to the same task. Ideas can thus merge, mix and cross-pollinate as the task actors involved collaborate on the work.
In such ways, human-computer co-creativity defines the computer system as equal collaborator in the creative process \cite{davis2013human}.

\section{Related Work}
The exploration of complex interactive workflows in CAC and MuMe has been fairly limited. 
This is partly because the rise of complex machine learning techniques such as deep learning have only recently been considered in questions of interaction research \cite{dove2017ux,yang2017role,yang2018machine,amershi2019guidelines}.
Within the current \textit{market-available} generative music applications \cite{melodrive,melodysauce,jukedeck,amperscore,spliqs,flowmachines,musenet}, and the \textit{academic systems} \cite{stylemachine,jnanalive,patter,folkrnn,impro-visor,pachet2003continuator,martin2011toolkit,maxwell2012manuscore,roberts2019magenta,tchemeube2019apollo,bougueng2022calliope} among others, few have been evaluated on their user experience, usability or acceptance. 
Bray et al. \cite{bray2017can} select three tools designed for CAC and analyze three specific computationally creative interface categories; direct manipulation systems, programmable interfaces and highly encapsulated systems; using respectively Jnana Live, Patter and StyleMachine systems. They conducted a preliminary investigation looking at the user experience of a single expert user using the tools and discussed the implications that encapsulating system’s functionalities has on visibility of user parameters across different computationally creative scenarios. 

Roberts et al. \cite{roberts2019magenta} develop Magenta Studio, a system bringing interactive generative music to professional music creators, combining deep learning-based music generation and direct integration with Ableton Live, a popular music software. Magenta Studio is a collection of five distinct music plugins for continuation, 4-bar generation, drum generation, interpolation, and rhythmic performance. 
The authors conduct a short survey using mixed questionnaire with early adopters (a mix of musicians, producers and machine learning enthusiasts) evaluating the system’s effectiveness and emerging experiences. From the 89 responses, they learn about the ease of using Magenta Studio's outputs in their musical work (66\% easy, 24\% neutral, 10\% difficult), the observed response time in producing generated outputs (58\% experienced little to no delay and only 6\% reported significant delays), the usefulness of the system in how easy it was to achieve desirable musical effect (41\% easy, 40\% neutral, 19\% difficult), whether the system made them feel more creative (72\% more, 20\% neutral, 8\% less), and more productive in their creative process (93\% yes).

Finally, Louie et al. \cite{louie2020novice} compare Bachdoodle \cite{huang2019bach}) with infill mask and the Cococo interface with interactive steering controls: voice lanes, semantic sliders, and generative alternatives. The within-study design of 21 novice musicians shows semantically-relevant user control features improves overall usability of the AI music generative model, namely on creative ownership (or authorship), self-efficacy and collaboration. The study measures 3 dimensions of user-AI interaction: the composition experience, the attitude towards AI and the perceptions of composition.
On composition experience, they record significant increase in expressing goals, self-efficacy, engagement and learning, with no significant difference on effort. On attitude towards AI, they observe an increase in controllability, comprehensibility, collaboration and trust. On perceptions of composition, they get increase in ownership, AI's vs user's contribution, completeness, and no significant difference on uniqueness of the generative outputs.
The study only addresses 4-voice piano-based polyphonic western music composition. Thus, it is not clear how the results might extend to other genres of music or multi-track generation.
 
Our study extends the literature by looking specifically at \textit{expert} composers, be it hobbyists and professionals, across many music genres (rock, pop, hip-hop, jazz, electronic), music of four instrument tracks and three \textit{distinct} music composition tasks. Our results complement existing findings while offering more granularity and providing a first account on technology acceptance for such creative AI music tools.

\section{Methodology}
For this study, our research 
is mixed methods, using both quantitative and qualitative research techniques.
Specifically, we adopt a triangular design \cite{creswell2003advanced} with a convergence model that arrives at interpretative results by comparing and contrasting our quantitative and qualitative data analysis.
The methodology for each of the research questions is outlined in the sub-sections below.

\subsection{Participants}
For this study, \textit{expert composers} are defined as composers with extensive experience with music composition, as well as experience with typical DAWs such as Cubase. The following two participant groups break down the expert population:

\begin{itemize}
    \item \textit{Hobbyist \textit{expert} composers (N=8)}: Music composition is not their main source of revenue. This group is more likely to engage openly with the software and music process. 
    \item \textit{Professional \textit{expert} composers (N=10)}: Music composition is a paid occupation and their main source of revenue. This group most likely has their own specialized compositional workflow.
\end{itemize}

\subsection{Interactive Interface}
MMM is integrated in Cubase's music workflow as the MMM-C system (Figure \ref{fig:mmmincubase}) and steerable via three sequential user actions: 

\begin{enumerate}
    \item The selection of \textit{bars} of symbolic music content, on an existing or new track. A musical bar is a unit of time measurement for symbolic music content and typically corresponds to the duration of 4 quarter notes of music (on a 4/4 time signature).
    \item The adjusting of the \textit{temperature} parameter (0-100\%, default: 50\%). Temperature is a common control parameter in generative neural network models. In the case of MMM, it controls how conservative (closer to 0\%) or experimental  (closer to 100\%) the generated musical content is, given what the model can typically generate.
    \item Then, the user clicks on the "Generate" button to trigger the generation of new musical sequences for the selected bars. MMM takes into account the vertical and horizontal context, and generates according to the selected instrument (e.g., for violin vs saxophone or piano, the model will generate stylistically different musical results). For this study and due to limitations when it comes to normalizing VST instrument types, the name of MIDI channels was enforced to match the instrument type and used as the input variable for the generation request.

\end{enumerate}

Given that the temperature parameter has been the most extensively used and the most commonly found parameter among modern generative music systems (due to the fact that models all share the property of stochasticity), we choose it as the main attribute control. This also helps relate our results to existing literature while expanding on other factors such as number of tracks, style agnosticity of the model, or user groups. Finally, because generation of new music by the system can take a few seconds, a "Cancel" button is available for the user to cancel the generative request. Undo-redo features already available in the Cubase software extends to the MMM-C, enabling the user to, if desired, revert generative changes made by the system.

\subsection{Tasks}
The participants are given three tasks  typical in multi-track music composition:

\begin{enumerate}
    \item \textbf{Arrangement} (\verb|ARR| / Task 1): Producing an arrangement of a 16-bar long composition. This consists in extending the composition with 3 new MIDI tracks with instrumentation.
    \item \textbf{Variation} (\verb|VAR| / Task 2): Producing a variation of an existing 16-bar long composition of their choice
    \item \textbf{Original Composition} (\verb|ORI| / Task 3): Generating a 16-bar long original piece of music given a seed MIDI file of their choosing as starting point.
\end{enumerate}

They are asked to complete the tasks by using MMM-C features of track in-filling, bar in-filling and temperature control. They are provided with existing genre-specific MIDI files to select from (pop, rock, electronic and classical). Alternatively, they can use their own MIDI work for the study. Participants are also allowed to freely edit the parts generated by MMM-C in order to achieve their tasks. 

\subsection{Measurement Tools}
We measure the following constructs:
\begin{enumerate}
    \item \verb|Usability|: the extent to which the system enables the user to effectively, efficiently and satisfiably achieve its goals.

    \item \verb|User experience|: the quality of the experience the user has when interacting with the system.

    \item \verb|Acceptance|: to understand the potential of adoption for such systems in the future.
\end{enumerate}

\subsubsection{Usability}
Our chosen evaluation technique for usability is remote unmoderated quantitative usability testing \cite{barnum2020usability}. This is because participants for this research are provided by Steinberg from their beta-testers pool of musicians and composers. 
The participants fill the following surveys:
\begin{itemize}
    \item The Standard System Usability Scale (SUS) (5-point Likert scale) \cite{brooke1996sus}  
    scores have a range of 0 to 100. A value $<$ 50 is considered \textit{unacceptable}, 50-70, \textit{marginal}, and $>$ 70, \textit{acceptable}.
    
    \item The Creativity Support Index (CSI) \cite{cherry2014quantifying} (12 agreement statements, 10-point Likert scale) measures the extent to which MMM-C effectively support the composer’s creative process. It outputs a single CSI score out of 100, with a higher score indicating greater creativity support. CSI's questions evaluated the following factors (2 statements per factor): Results Worth Effort, Exploration, Collaboration, Immersion, Expressiveness and Enjoyment. 
    Our CSI measurements are adjusted to use one statement per factor (5-point Likert scale) and because our MMM-C did not involve any collaboration, we omit the corresponding, ending up with a total of 5 agreement statements. 

    \item Controllability: For measuring the tool's functional controllability or AI-steerability, we use a custom set of two 10-point Likert scale questions and an open-text comment for capturing complementary qualitative details to the scale questions. 
\end{itemize}

\subsubsection{User Experience}
We are interested in user feedback about the software’s capabilities as well as in understanding the level of:
\begin{itemize}
    \item Trust and perceived quality: regarding of the system’s actions and the generated musical outputs
    \item Perceived authorship: the degree to which the participants consider themselves authors of the final music piece.
    \item Flexibility: the extent to which the MMM model is flexible when assisting the participant in achieving compositional objectives. 
\end{itemize}

Thus, each participant completes
    four open-ended qualitative questions about their feeling of trust, perceived quality, authorship and flexibility. Additionally, they answer
    eight reflective questions regarding their experience and the major benefits and inconveniences associated with music creation using a creative AI tool like MMM.
We apply inductive/open coding to identify resulting themes from the data.


\subsubsection{Acceptance}
The Technology Acceptance Model (TAM) \cite{davis1989perceived} is used to evaluate the acceptance level from the participants by measured three factors: perceived usefulness, perceived ease of use, and attitudes towards usage of the system. 
Our TAM questionnaire consists of 12 questions (5-point Likert scale) with half (6) measuring perceived ease of use and the other half, the participant's perceived usefulness.
Scores for individual questions are averaged out to obtain a single score for each of 1) \textit{perceived ease of use} and 2) \textit{perceived usefulness}.
TAM has been shown to reliably predict and explain user acceptance of information technologies \cite{davis1996critical} and is the most widely used instrument for this purpose. 



\section{Study}
\label{Study}

\begin{figure*}
\centering
\includegraphics[width=.85\linewidth]{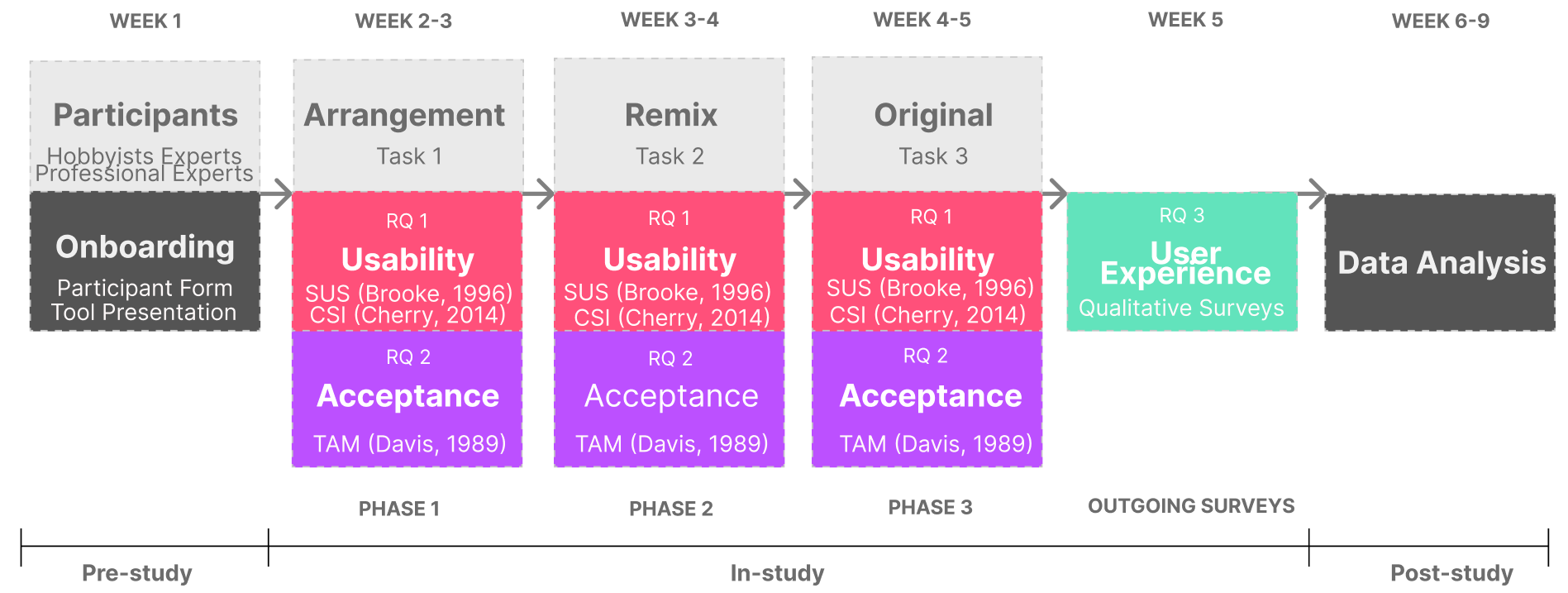}
\caption{MMM-C Interaction Study Plan \& Process}
\label{fig:researchprocess}
\end{figure*}

The study was organized as depicted in Figure \ref{fig:researchprocess}.
We ran a pre-study mockup with internal participants (i.e., lab members) to test our protocol and review procedures. From this, we updated research materials, participant instructions and surveys to refine the study process. We then recruit participants via Steinberg's pool of beta-testers. These testers are expert users of various Steinberg software including Cubase and volunteered to be part of the program. Ethics approval (\textit{ID \#30000223}) was obtained from the SFU's Research Ethics Board to conduct this study. 
A \textit{Participant Guide} document provided to all participants includes checklist of research activities to be completed with links to relevant resources.
The entire study run takes between 20h to 30h of effort for a given participant depending on their level of familiarity with such creative AI systems, Cubase and research processes. The participant's effort is spread over a four to five weeks time period. The study run from June 15th to July 25th 2022. Participants were asked to fill in an onboarding form. 
They went through all three tasks and all four survey phases. They then uploaded their results (project file, an audio master, and usage log) to the Steinberg Share cloud portal. Participants were compensated with a choice among Steinberg's music software licenses such as Cubase Element [Audio Production] (99.99 EUR), Dorico Element [Scorewriting] (99.99 EUR), Wavelab Element [Audio Mastering] (129.00 EUR) or 1-year time limited license of Cubase Pro [Audio Production] (approx. 124.85 EUR). 

\section{Results}

Out of the 34 participants who completed the onboarding form, a total of 18 actually took part in the study by completing at least one of the study tasks. The next sub-sections detail our results. An anonymized sample collection of the musical excerpts by participants can be found at \href{https://soundcloud.com/mmmcubase/sets/mmm-cubase?si=8cfe6eda0a1f4d0ba59f170cd8dd6ab1}{\texttt{soundcloud.com/mmmcubase/sets/mmm-cubase}}. 

\subsection{Demographics}
The participant distribution is balanced with 8 hobbyist experts and 10 professional experts in the study. 17/18 identify as a "male" and 1/18 "prefers not to say". Age groups breakdown as follow: 20-29 years old (1/18), 30-39 years old (5/18), 40-49 years old (5/18), 50-59 years old (5/18), $\geq$ 60 years old (2/18). 
     \textbf{16/18} participants identify as "\textbf{Expert User}". Although, we were mainly concerned with expert composers, we had a minority of other user profiles, namely "Enthusiasts" (2/18).
     Participants' geographic locations show a diverse spread: Australia = 1, Brazil = 2, France = 1, Germany = 2, Netherlands = 1, Portugal = 1, South Africa = 1, Spain = 1, USA: 3, UK = 2 and Canada = 3.
     All participants (18/18) report being familiar with at least one of the Cubase or Nuendo \footnote{Nuendo is a DAW used for post-production in film and video games with advanced sound design, dialog recording and game audio features.} software.   
     16/18 reported using MIDI or Instrument tracks (symbolic composition) in Cubase/Nuendo.
     Only 7/18 have prior experience with music-related AI systems while the rest (11/18) does not.
     The \textbf{average musical experience level} is \textbf{8.1} /10 (SD = 1.43); with the hobbyists group at 7.13 (SD = 1.17) and the professionals group at 8.80 (SD = 1.17).
     Similarly, the average experience level with Cubase/Nuendo is 6.89 /10 (SD = 2.9); with the hobbyists group at 6.63 (SD = 2.39) and the professionals group at 7.10 (SD = 3.24).

\subsection{Usability}
Overall, the quantitative results indicates that MMM-C is easy to use and to operate. However, participants struggle with steering the tool to produce their desired outcome.

\subsubsection{Standard Usability Scores (SUS)}
\begin{figure*}
\includegraphics[width=.85\linewidth]{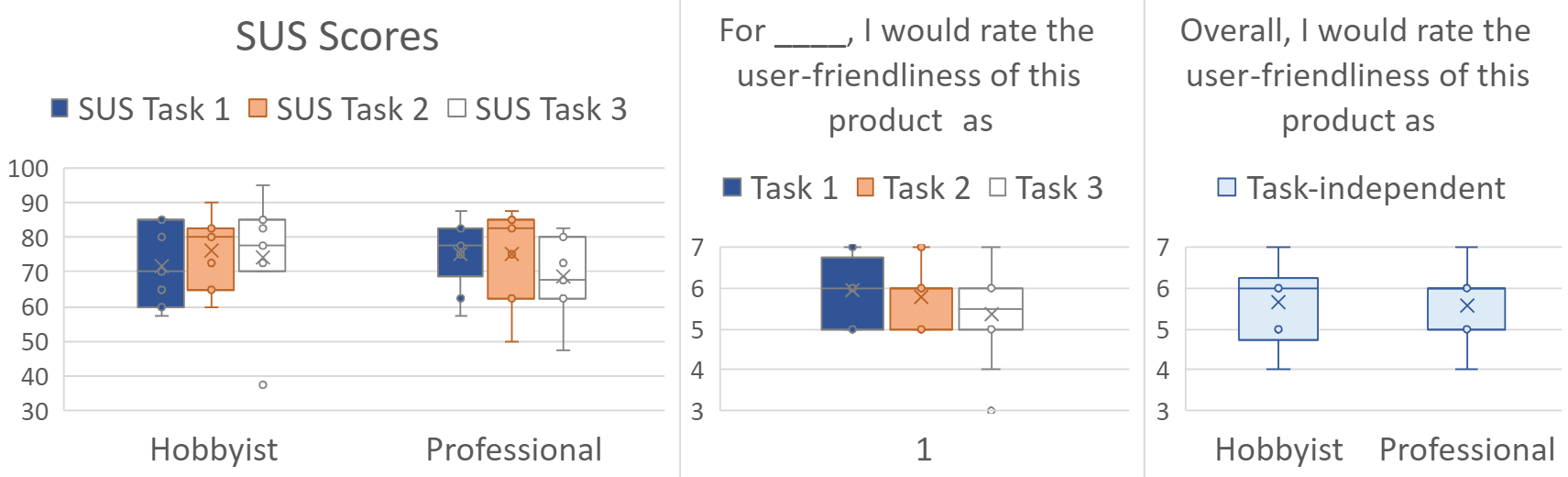}
\centering
\caption{Task-based SUS Results + Task-based and Overall User-Friendliness Scores}
\label{fig:combined_sus}
\end{figure*}

SUS Task 1 is 73.75 / 100 (SD = 10) , SUS Task 2 is 75.71 (SD = 11.59) and SUS Task 3 is 71.43 (SD = 14.48). Overall, SUS scores for MMM-C across tasks are \textit{acceptable}. 
Figure \ref{fig:combined_sus} reports a breakdown of SUS scores between hobbyists and professionals across tasks.
Participants also rated the user-friendliness of the tool, which is a one-value rating score given after each task and at the end of the study (Figure \ref{fig:combined_sus}) \footnote{For all presented graphs, box edges are 25th and 75th percentiles, the cut line is the median, "x" is the mean, the circles are inner datapoints, and the lower and upper wiskers are the minimum and maximum datapoints.}. Generally, it indicates that the system is easy to use for both groups and make sense given the 1-parameter interface design approach. We did not find any significant difference between SUS values across tasks and user-friendliness scores between hobbyist and professional groups; except for the task-based  friendliness scores between task 1 (\verb|ARR|) (5.94 / 10, SD = 0.75) and task 3 (\verb|ORI|) (5.36, SD = 0.97). This result indicates that MMM-C is more user-friendly for task 1 compared to task 3 (p=0.03). Qualitative data seems to hint at this; P22 writes, with respect to comments on controllability "\textit{to get a sound that suited the instrument or music style was not easy - especially when composing my own rather than the first two projects}".

\subsubsection{Controllability}
\begin{figure}
\includegraphics[width=.9\linewidth]{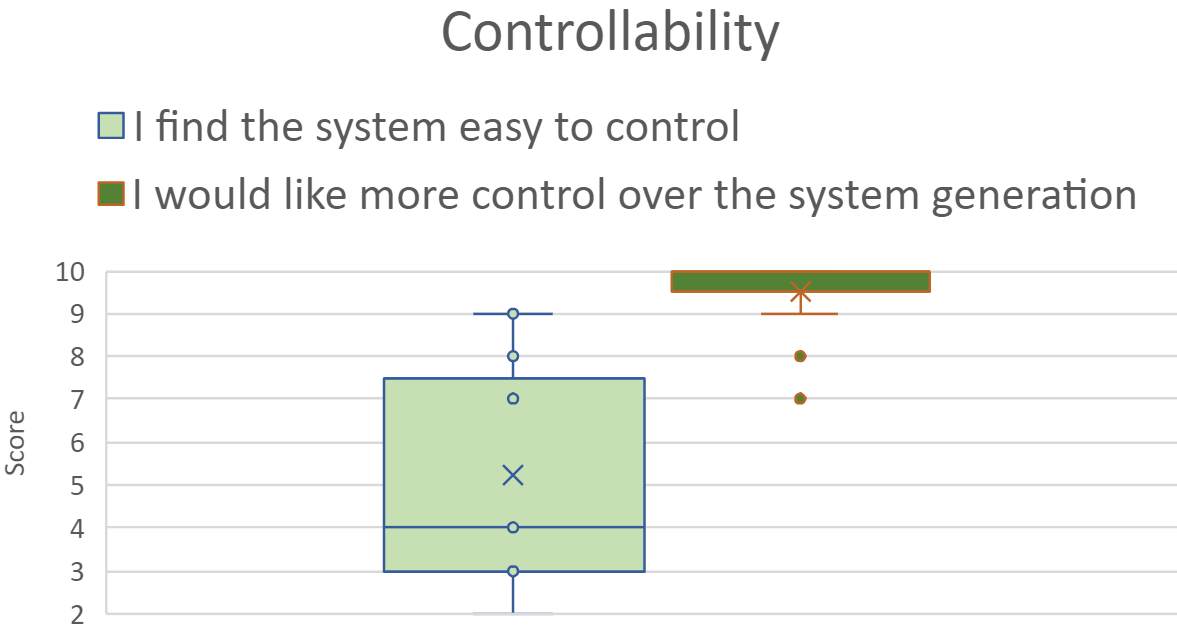}
\caption{MMM-Cubase's Controllability Scores}
\label{fig:ovr_controllability}
\end{figure}
Controllability scores are measured on two dimensions measured at the end of the study: ease of control over the system (5.23 / 10, SD = 2.52) and whether more control is desirable (9.54, SD = 0.93). These are later explained and shown to be consistent with observations from qualitative data. There were 4 main sub-themes that emerged: "Difficulty steering the system" (9/18), "easy to use" (4/18), "lack of parameters" (4/18), "want more flexibility" (6/18). Participants encountered challenges with steering the system while anticipating additional controls that could help them with the tasks. P1 writes "\textit{It seemed hard to understand the actual effects of the single control for the system. As such it was hard to feel in control of the result. However, using the system experimentally without specific results in mind was enjoyable.}". 
Participants also report difficulty relying solely on the \textit{temperature} parameter to drive generation. 
The "easy to use" sub-theme shows that MMM-C's interface was easy to use; though we can also notice that it is intertwined with the issue of lack of parameters. For example, P4 answers "\textit{The system is easy to control from a interface perspective. However, the lack of any type of parameters makes it feel like a more random process than a creative one.}" while P21 writes \textit{"I understand the idea was to create the simplest GUI for the user and the tool is super easy to use"}.
Finally, on the theme of "want more flexibility", participants suggest interactions that they would prefer to achieve their user goals. 
P18 says "\textit{Enter a phrase or/and rhythms to drive the system}" and P22 writes "\textit{For controllability, I would like to see suggestions for instruments in terms of 'this will suit a bass sound' or 'this will suit funk genre of music' -for example.}". 

\subsubsection{Creativity Support Index}
\begin{figure}
\includegraphics[width=1\linewidth]{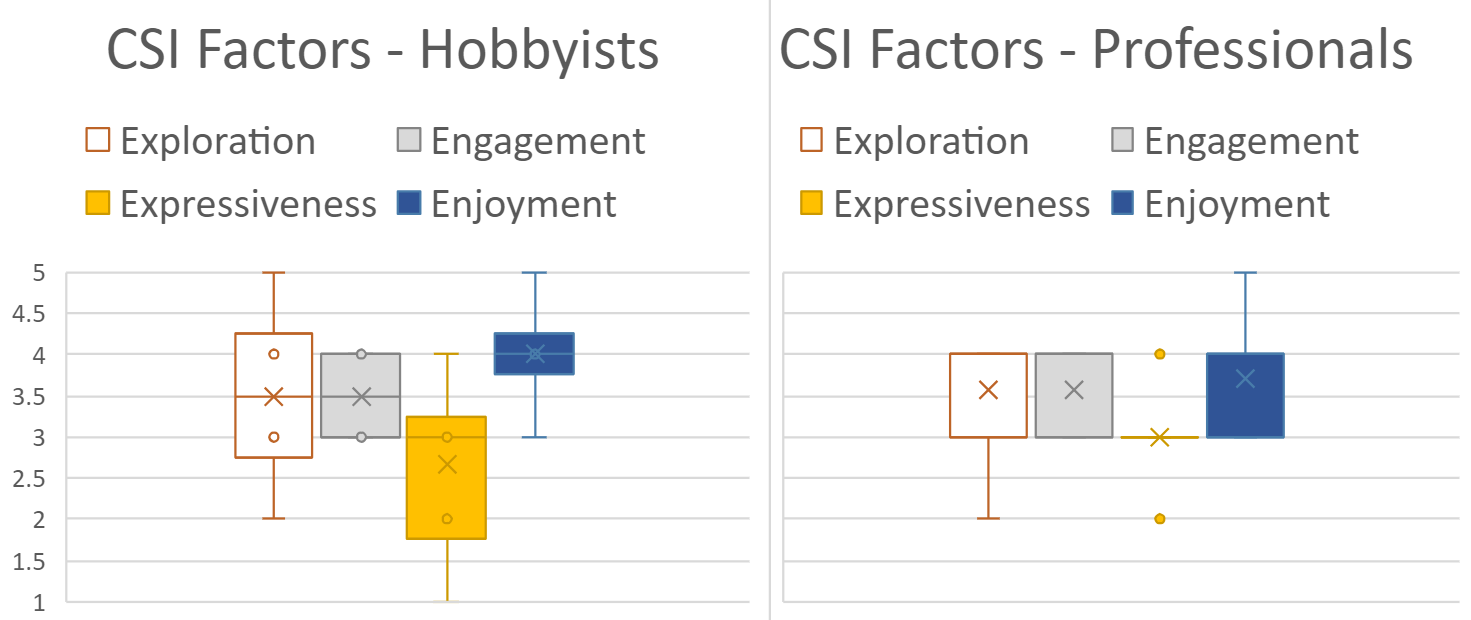}
\caption{CSI Factor Scores per Participant Group}
\label{fig:combined_csi_scores}
\end{figure}
CSI ratings were completed by 13 out of the 18 participants. CSI factors give us a glimpse into the participant experience and help us understand what the tool can and cannot do with respect to creativity support:
\begin{itemize}
    \item Exploration is fairly good (3.54 / 5, SD = 0.84) despite some outlier data which suggests otherwise.
    \item Enjoyment is the highest metrical score (3.85, SD = 0.66) and is consistent across participants. This is consistent with the theme in our qualitative data about users enjoying using the system and multiple experiences of positive surprises on musical results/ideas using the tool.
    \item Expressiveness is the least performing metric (2.85, SD = 0.77). This is consistent with and corroborates qualitative reports of frustation experienced by users due to the inability to control/steer the system with just 1-parameter and limited understanding of the system's behavior/mechanics. This is consistent and validates previous literature \cite{louie2020novice,bray2017can} on this topic and the necessity of careful parameter design for non-deterministic music AI systems. Qualitative data suggest that this metric could be improved with increased user control or system's capability on guiding musical outputs e.g., by exposing more of MMM's extensive control parameters into the MMM-C system's interface (e.g., duration controls, polyphony controls, note density).
    \item Engagement is fairly good (3.54, SD = 0.49) across participants. 
\end{itemize}

The Figure \ref{fig:combined_csi_scores}  presents a breakdown of CSI Factors between hobbyists and professionals.

\subsection{User Experience}
For our coding strategy, we had two coders apply inductive/open coding separately and independently to each survey question, then later discussed and consolidate them into a set of reliable codes. One coder primarily focused on identifying cross-question emerging \textit{themes}, and the other focused on identifying \textit{themes} within each question. Table~\ref{tab:CodesByQuestion} shows the themes as they occur in each question. For example, in the question relating to \textbf{Trust}, participants discuss \textit{repeated generations}, \textit{non-determinism}, and \textit{co-creation}; while co-creation is also brought up by participants in questions on \textbf{Authorship}, \textbf{Workflow} and \textbf{Practical Benefits}. 
We identified the following broad themes:

\begin{table}[]
\parbox{\linewidth}{
\centering
\resizebox{0.477\textwidth}{!}{%
\begin{tabular}{l|l}
\toprule
\Large \textbf{Question} & \Large \textbf{Observed Themes} \\

\midrule
\Large Trust & \Large Repeated generation,\\ & \Large Non-determinism, Co-creation \\
\Large Authorship  & \Large Co-creation, Creative control,\\ & \Large Creative ownership  \\
\Large Flexibility  & \Large Parametric control, Creative control    \\
\Large Challenges   & \Large System knowledge, Repeated generation,\\ & \Large Non-determinism, Creative control\\
\Large Features  & \Large Parametric control, Ease of use\\
\Large Workflow & \Large Surprise \& Novelty, Co-creation\\
\Large Concerns & \Large Surprise \& Novelty, Co-creation \\
\Large Practical benefits & \Large Non-determinism, Surprise \& Novelty,\\ & \Large Co-creation\\
\Large Quality of & \Large Repeated generation, Co-creation\\ \Large generated music & \\ 
\Large Future use & \Large Surprise \& Novelty, Co-creation\\
\Large Releasable & \Large Parametric control, Surprise \& Novelty \\
\Large Final thoughts & \Large Parametric control\\                                                                                                     \bottomrule  
\end{tabular}
}
\caption{List of questions and the themes that appear}
\label{tab:CodesByQuestion}
}
\end{table}

\begin{itemize}
    \item \textbf{Parametric control}
Users discuss the explicit parameters used to control output in MMM-C. Participants discuss how single parameter allows for rapid iteration and easy interaction. Participants mention the limitations of having a single parameter to manipulate. Participants also describe confusion in terms of how the controllable parameter affects the musical output and propose multiple suggestions for additional parameters.

    \item \textbf{Ease of Use}
Participants discuss how MMM-C is easy to operate and integrate into their process. While participants discuss challenges in using the system to accomplish a particular goal (see “Parametric control” and “Creative control”), they note that the interaction itself is easy.

    \item \textbf{Repeated generation}
Participants regularly mention generating music multiple times to find a suitable output. Most commonly, participants describe feeling the need to generate many times in order to find a suitable output.

    \item \textbf{Non-determinism}
Participants describe the unpredictable nature of MMM-C’s output. This code is closely related to the repeated generation code, as participants often describe unpredictability in output as a reason for repeated generations.

    \item \textbf{Surprise \& Novelty}
Participants describe both how MMM-C’s output is surprising in general, and how MMM-C generates variations that are not what the user would have themselves written. This is a particularly interesting finding given that \textit{surprise} and \textit{novelty}, along with \textit{value}, arguably represent the three constructs used to define creativity \cite{grace2015data}.

    \item \textbf{Co-creation}
Participants describe curating and editing the output of MMM-C to achieve their creative vision. Participants mainly describe using MMM-C as a source of inspiration or new ideas. Participants note that the non-deterministic behavior of MMM-C leads to heavy curation of the output, and that they generally edit on the generated content.

    \item \textbf{Creative control}
Participants describe how much they feel able to control their final compositions. This code includes similar elements to the “parametric control”, but encompasses the participant’s own musical input.

    \item \textbf{Creative ownership}
Participants describe how much ownership they feel over the final creative output. Participants describe some anxiety about the use of “AI” tools, or feel that the tool provides the creativity. In general, participants feel that they still have creative ownership of the final output
\end{itemize}


Few final points highlight that interacting with the system helped them see the potential of such CAC systems. P4 writes "\textit{Overall using the MMM in this context helped me see the potential of such systems in the creative process.}", P34 reports "\textit{I see the potential and love to participate.}" while P31 says "\textit{I really enjoyed participating in this, and would love to be invited to participate in future iterations of this to see how this evolves - I can see that this has plenty of potential!}". Also, in reference to the vast number of papers that raise fear of work replacement, the users did not express such concerns in the study.

\subsection{Acceptance}
TAM results are as follow: 
\begin{itemize}
    \item Task 1: Perceived Ease of Use = 3.12 / 5 (SD = 0.91), Perceived Usefulness= 3.56 (SD = 0.51)
    \item Task 2: Perceived Ease of Use = 3.26 (SD = 0.89), Perceived Usefulness = 3.68 (SD = 0.63)
    \item Task 3: Perceived Ease of Use = 3.14 (SD = 0.96), Perceived Usefulness = 3.42 (SD = 0.62)
\end{itemize}

\begin{figure*}
\includegraphics[width=\linewidth]{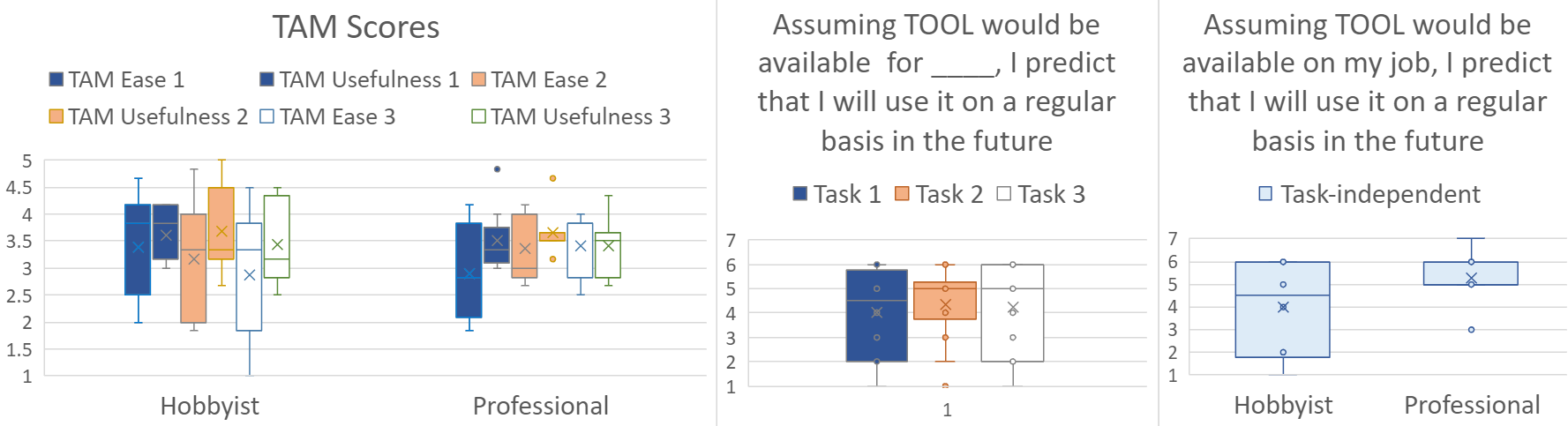}
\caption{Task-based TAM Results + Task-based and Overall 1-Value Rating TAM Scores}
\label{fig:combined_tam_scores}
\end{figure*}

In general, participant scores on future predicted use of the MMM-C tool are above average, however not significant enough to be conclusive. This is somewhat of an expected result given that we purposefully limit their affordance of the system to characterize behavior.
Figure \ref{fig:combined_tam_scores} reports a breakdown of TAM scores between hobbyists and professionals across tasks.
Participants also rated their likelihood of regularly using the tool by reporting one-value scores on TAM per task and overall.
Finally, Figure \ref{fig:combined_tam_scores} also indicates that acceptance scores at task-level are consistent with the final one-value rating scores from the outgoing surveys (filled out at the end of the study). We did not find any significant difference between TAM values across tasks and one-value rating scores between hobbyist
and professional groups.
Overall, acceptance results indicate that more must be done about such systems before they are considered for "serious" musical use cases, primarily on flexibility and exposing additional controls to help the user achieve their creative goals.

\section{Discussion}
Overall usability and acceptance levels are positive. Though, results indicate, mainly through qualitative data and controllability scores, that a 1-parameter design is not enough for generative music co-creation in the case of expert composers, even for an expressive model like MMM. This is not necessarily evident since parameter design is at least dependent on the expressive capacity of the generative model and the quality level of creative outputs required by the user, thus dependent on the target user profile.
The possible variations in the standard of acceptable outputs between experts, novices or other user groups should be considered when  evaluating parameter design for co-creative AI systems.
Finally, we believe that our results are generalisable to any DAW since Cubase offers typical features, music environment and workflows comparable to other standard commercial DAWs. 
Our proposed methodological assemblage is also relevant for other generative tasks such as language or visual in-painting. 
Finally, with respect to our initial research questions on MMM, we find that, given a system like MMM-C which greatly reduces MMM's generative power and controls, the model is still usable and relatively effective at generating musical content that helps advance experts' creative process. We gain insight into issues of autonomy, controllability and authorship where opening up the interface to MMM's attribute controls (e.g., note density, polyphony range, style control) could significantly enhance the musician's creative experience. We find that the level of technology acceptance is positive overall and we expect this metric to increase with a better controllable interface.

\section{Conclusion and Future Work}
We presented an evaluation of MMM-C, a co-creative AI tool for CAC with promising results. We conducted an experiment, contributing an original methodological design, measuring the user-experience, usability and technology acceptance of MMM-C with both groups of expert-level hobbyist and professional composers. The results show that, although a basic 1-parameter minimal interface design to steer a non-deterministic music generative system can be problematic for experts, it can still offer valuable opportunity for exploration of novel creative ideas and help them address writer's block in music composition. We also found that such use cases of \textit{out-of-the-box unfocused exploration} can bring an enjoyable experience. As an illustration, P30 answers in the "Final Thoughts" question: \textit{"This gave me back the joy of composing for the sake of it. Thanks!"}. 

Beyond the scope of this paper, we intend to re-run our methodology protocol on an interface which exposes to the user, more of MMM's available controls (e.g., note polyphony, duration and/or density controls), looking to understand more complex parametric interactions. This includes a fully functional interface integrating the model into the Cubase MIDI editor. 
The promises of co-creative interfaces are 1) an increase of the efficiency of professional musicians to complete musical tasks and 2) a lower bar of entry for beginner musicians to express musical ideas. On the later, adapting our methodology and co-creative system to study beginner composers, for example, by running an experiment using freely accessible DAW, can be achieved and contrasted against findings on expert composers.
As the integration of CAC systems in professional music ecosystems continues, answers to our larger research questions could better inform interface design for such systems and characterize more precisely the nature of human-AI co-creation. This is particularly timely in an era where the non-deterministic behavior of interactive systems powered by machine learning techniques such as neural networks and deep learning poses new research challenges.



\section*{Acknowledgments}

We would like to acknowledge the canadian funding bodies Mitacs and the Social Sciences and Humanities Research Council (SSHRC) for making this research possible via, respectively, the Mitacs Accelerate and SSHRC Partnership grant programs. We also would like to thank Steinberg Media Technologies GmbH as our research partner for this study, and Alex Kitson for her valuable feedback on research design.

\bibliographystyle{named}
\bibliography{ijcai23}

\begin{thebibliography}{}

\bibitem[\protect\citeauthoryear{Alemi and Pasquier}{2017}]{alemi2017walknet}
Omid Alemi and Philippe Pasquier.
\newblock Walknet: A neural-network-based interactive walking controller.
\newblock In {\em International Conference on Intelligent Virtual Agents},
  pages 15--24. Springer, 2017.

\bibitem[\protect\citeauthoryear{Alemi \bgroup \em et al.\egroup
  }{2017}]{alemi2017groovenet}
Omid Alemi, Jules Fran{\c{c}}oise, and Philippe Pasquier.
\newblock Groovenet: Real-time music-driven dance movement generation using
  artificial neural networks.
\newblock {\em networks}, 8(17):26, 2017.

\bibitem[\protect\citeauthoryear{Amershi \bgroup \em et al.\egroup
  }{2019}]{amershi2019guidelines}
Saleema Amershi, Dan Weld, Mihaela Vorvoreanu, Adam Fourney, Besmira Nushi,
  Penny Collisson, Jina Suh, Shamsi Iqbal, Paul~N. Bennett, Kori Inkpen, Jaime
  Teevan, Ruth Kikin-Gil, and Eric Horvitz.
\newblock Guidelines for human-ai interaction.
\newblock In {\em Proceedings of the 2019 CHI Conference on Human Factors in
  Computing Systems}, CHI '19, page 1–13, New York, NY, USA, 2019.
  Association for Computing Machinery.

\bibitem[\protect\citeauthoryear{Anderson \bgroup \em et al.\egroup
  }{2013}]{stylemachine}
Christopher Anderson, Arne Eigenfeldt, and Philippe Pasquier.
\newblock The generative electronic dance music algorithmic system
  ({G}{E}{D}{M}{A}{S}).
\newblock In {\em Proceedings of the AAAI Conference on Artificial Intelligence
  and Interactive Digital Entertainment}, volume~9, pages 5--8, 2013.

\bibitem[\protect\citeauthoryear{Barnum}{2020}]{barnum2020usability}
Carol~M Barnum.
\newblock {\em Usability testing essentials: Ready, set... test!}
\newblock Morgan Kaufmann, 2020.

\bibitem[\protect\citeauthoryear{Bray \bgroup \em et al.\egroup
  }{2017}]{bray2017can}
Liam Bray, Oliver Bown, and Benjamin Carey.
\newblock How can we deal with the design principle of visibility in highly
  encapsulated computationally creative systems?
\newblock In {\em Eighth International Conference on Computational Creativity,
  ICCC, Atlanta}, pages 65--71, 2017.

\bibitem[\protect\citeauthoryear{Briot \bgroup \em et al.\egroup
  }{2017}]{briot2017deep}
Jean-Pierre Briot, Ga{\"e}tan Hadjeres, and Fran{\c{c}}ois Pachet.
\newblock Deep learning techniques for music generation -- a survey.
\newblock {\em arXiv preprint:1709.01620}, 2017.

\bibitem[\protect\citeauthoryear{Brooke}{1996}]{brooke1996sus}
John Brooke.
\newblock Sus: A 'quick and dirty' usability scale.
\newblock {\em Usability evaluation in industry}, 189(3), 1996.

\bibitem[\protect\citeauthoryear{Cherry and
  Latulipe}{2014}]{cherry2014quantifying}
Erin Cherry and Celine Latulipe.
\newblock Quantifying the creativity support of digital tools through the
  creativity support index.
\newblock {\em ACM Transactions on Computer-Human Interaction (TOCHI)},
  21(4):1--25, 2014.

\bibitem[\protect\citeauthoryear{Clark \bgroup \em et al.\egroup
  }{2018}]{clark2018creative}
Elizabeth Clark, Anne~Spencer Ross, Chenhao Tan, Yangfeng Ji, and Noah~A.
  Smith.
\newblock Creative writing with a machine in the loop: Case studies on slogans
  and stories.
\newblock In {\em 23rd International Conference on Intelligent User
  Interfaces}, IUI '18, page 329–340, New York, NY, USA, 2018. Association
  for Computing Machinery.

\bibitem[\protect\citeauthoryear{Creswell \bgroup \em et al.\egroup
  }{2003}]{creswell2003advanced}
John~W Creswell, Vicki~L Plano~Clark, Michelle~L Gutmann, and William~E Hanson.
\newblock {\em Advanced mixed methods research designs}, pages 209--240.
\newblock Sage, 2003.

\bibitem[\protect\citeauthoryear{CSL}{2013}]{flowmachines}
Sony CSL.
\newblock Flowmachines.
\newblock \url{https://flow-machines.com/}, 2013.
\newblock Accessed: 2022-05-25.

\bibitem[\protect\citeauthoryear{Davis and Venkatesh}{1996}]{davis1996critical}
Fred~D Davis and Viswanath Venkatesh.
\newblock A critical assessment of potential measurement biases in the
  technology acceptance model: Three experiments.
\newblock volume~45, pages 19--45, USA, jul 1996. Academic Press, Inc.

\bibitem[\protect\citeauthoryear{Davis \bgroup \em et al.\egroup
  }{2016}]{davis2016empirically}
Nicholas Davis, Chih-PIn Hsiao, Kunwar Yashraj~Singh, Lisa Li, and Brian
  Magerko.
\newblock Empirically studying participatory sense-making in abstract drawing
  with a co-creative cognitive agent.
\newblock In {\em Proceedings of the 21st International Conference on
  Intelligent User Interfaces}, IUI '16, page 196–207, New York, NY, USA,
  2016. Association for Computing Machinery.

\bibitem[\protect\citeauthoryear{Davis}{1989}]{davis1989perceived}
Fred~D Davis.
\newblock Perceived usefulness, perceived ease of use, and user acceptance of
  information technology.
\newblock {\em MIS quarterly}, pages 319--340, 1989.

\bibitem[\protect\citeauthoryear{Davis}{2013}]{davis2013human}
Nicholas Davis.
\newblock Human-computer co-creativity: Blending human and computational
  creativity.
\newblock In {\em Proceedings of the AAAI Conference on Artificial Intelligence
  and Interactive Digital Entertainment}, volume~9, pages 9--12, 2013.

\bibitem[\protect\citeauthoryear{Donahue \bgroup \em et al.\egroup
  }{2019}]{donahue2019lakhnes}
Chris Donahue, Huanru~Henry Mao, Yiting~Ethan Li, Garrison~W Cottrell, and
  Julian McAuley.
\newblock Lakhnes: Improving multi-instrumental music generation with
  cross-domain pre-training.
\newblock {\em arXiv preprint:1907.04868}, 2019.

\bibitem[\protect\citeauthoryear{Dove \bgroup \em et al.\egroup
  }{2017}]{dove2017ux}
Graham Dove, Kim Halskov, Jodi Forlizzi, and John Zimmerman.
\newblock U{X} design innovation: Challenges for working with machine learning
  as a design material.
\newblock In {\em CHI Conference on Human Factors in Computing Systems}, CHI
  '17, page 278–288, New York, NY, USA, 2017. Association for Computing
  Machinery.

\bibitem[\protect\citeauthoryear{Ens and Pasquier}{2020a}]{ens2020flexible}
Jeff Ens and Philippe Pasquier.
\newblock Flexible generation with the multi-track music machine.
\newblock In {\em the 21st International Society for Music Information
  Retrieval Conference, ISMIR}, 2020.

\bibitem[\protect\citeauthoryear{Ens and Pasquier}{2020b}]{ens2020mmm}
Jeff Ens and Philippe Pasquier.
\newblock M{M}{M}: Exploring conditional multi-track music generation with the
  transformer.
\newblock {\em arXiv preprint:2008.06048}, 2020.

\bibitem[\protect\citeauthoryear{Ens and Pasquier}{2021}]{ens2021building}
Jeffrey Ens and Philippe Pasquier.
\newblock Building the {M}eta{M}{I}{D}{I} dataset: Linking symbolic and audio
  musical data.
\newblock In {\em ISMIR}, pages 182--188, 2021.

\bibitem[\protect\citeauthoryear{Evabeat}{2019}]{melodysauce}
Evabeat.
\newblock Melodysauce.
\newblock \url{https://evabeat.com/}, 2019.
\newblock Accessed: 2022-05-25.

\bibitem[\protect\citeauthoryear{Florin}{2015}]{patter}
Adam Florin.
\newblock Patter.
\newblock \url{https://cycling74.com/projects/patter-1}, 2015.
\newblock Accessed: 2022-05-25.

\bibitem[\protect\citeauthoryear{Grace \bgroup \em et al.\egroup
  }{2015}]{grace2015data}
Kazjon Grace, Mary~Lou Maher, Douglas Fisher, and Katherine Brady.
\newblock Data-intensive evaluation of design creativity using novelty, value,
  and surprise.
\newblock {\em International Journal of Design Creativity and Innovation},
  3(3-4):125--147, 2015.

\bibitem[\protect\citeauthoryear{Guzdial \bgroup \em et al.\egroup
  }{2019}]{guzdial2019friend}
Matthew Guzdial, Nicholas Liao, Jonathan Chen, Shao-Yu Chen, Shukan Shah,
  Vishwa Shah, Joshua Reno, Gillian Smith, and Mark~O Riedl.
\newblock Friend, collaborator, student, manager: How design of an ai-driven
  game level editor affects creators.
\newblock In {\em CHI Conference on Human Factors in Computing Systems}, CHI
  '19, page 1–13, New York, NY, USA, 2019. Association for Computing
  Machinery.

\bibitem[\protect\citeauthoryear{Huang \bgroup \em et al.\egroup
  }{2018}]{huang2018music}
Cheng-Zhi~Anna Huang, Ashish Vaswani, Jakob Uszkoreit, Ian Simon, Curtis
  Hawthorne, Noam Shazeer, Andrew~M Dai, Matthew~D Hoffman, Monica Dinculescu,
  and Douglas Eck.
\newblock Music transformer: Generating music with long-term structure.
\newblock In {\em International Conference on Learning Representations}, 2018.

\bibitem[\protect\citeauthoryear{Huang \bgroup \em et al.\egroup
  }{2019}]{huang2019bach}
Cheng-Zhi~Anna Huang, Curtis Hawthorne, Adam Roberts, Monica Dinculescu, James
  Wexler, Leon Hong, and Jacob Howcroft.
\newblock The bach doodle: Approachable music composition with machine learning
  at scale.
\newblock {\em arXiv preprint:1907.06637}, 2019.

\bibitem[\protect\citeauthoryear{Kantosalo \bgroup \em et al.\egroup
  }{2015}]{kantosalo2015interaction}
Anna~Aurora Kantosalo, Jukka~Mikael Toivanen, Hannu Tauno~Tapani Toivonen,
  et~al.
\newblock Interaction evaluation for human-computer co-creativity: A case
  study.
\newblock In {\em Proceedings of the sixth international conference on
  computational creativity}. Brigham Young University, 2015.

\bibitem[\protect\citeauthoryear{Keller}{2008}]{impro-visor}
Robert~M. Keller.
\newblock Impro-visor.
\newblock \url{https://www.cs.hmc.edu/~keller/jazz/improvisor/}, 2008.
\newblock Accessed: 2022-05-25.

\bibitem[\protect\citeauthoryear{Kranabetter \bgroup \em et al.\egroup
  }{2022}]{kranabetter2022audio}
Joshua Kranabetter, Craig Carpenter, Renaud~Bougueng Tchemeube, Philippe
  Pasquier, and Miles Thorogood.
\newblock Audio metaphor 2.0: An improved classification and segmentation
  pipeline for generative sound design systems.
\newblock In {\em Sound and Music Computing Conference (SMC)}, 2022.

\bibitem[\protect\citeauthoryear{Liang \bgroup \em et al.\egroup
  }{2017}]{liang2017automatic}
Feynman~T Liang, Mark Gotham, Matthew Johnson, and Jamie Shotton.
\newblock Automatic stylistic composition of bach chorales with deep lstm.
\newblock In {\em ISMIR}, pages 449--456, 2017.

\bibitem[\protect\citeauthoryear{Louie \bgroup \em et al.\egroup
  }{2020}]{louie2020novice}
Ryan Louie, Andy Coenen, Cheng~Zhi Huang, Michael Terry, and Carrie~J Cai.
\newblock Novice-{A}{I} music co-creation via {A}{I}-steering tools for deep
  generative models.
\newblock In {\em Proceedings of the 2020 CHI conference on human factors in
  computing systems}, pages 1--13, 2020.

\bibitem[\protect\citeauthoryear{Martin \bgroup \em et al.\egroup
  }{2011}]{martin2011toolkit}
Aengus Martin, Craig~T Jin, and Oliver Bown.
\newblock A toolkit for designing interactive musical agents.
\newblock In {\em Proceedings of the 23rd Australian Computer-Human Interaction
  Conference}, pages 194--197. ACM, 2011.

\bibitem[\protect\citeauthoryear{Maxwell \bgroup \em et al.\egroup
  }{2012}]{maxwell2012manuscore}
James Maxwell, Arne Eigenfeldt, and Philippe Pasquier.
\newblock Manuscore: Music notation-based computer assisted composition.
\newblock In {\em ICMC 2012: Non-Cochlear Sound - Proceedings of the
  International Computer Music Conference}, 2012.

\bibitem[\protect\citeauthoryear{Maxwell \bgroup \em et al.\egroup
  }{2016}]{spliqs}
James Maxwell, Nicolas Gonzalez~Thomas, and Olivier Vincent.
\newblock Spliqs.
\newblock \url{https://spliqs.com/}, 2016.
\newblock Accessed: 2022-05-25.

\bibitem[\protect\citeauthoryear{Music}{2014}]{amperscore}
Amper Music.
\newblock Amperscore.
\newblock \url{https://ampermusic.com/}, 2014.
\newblock Accessed: 2022-05-25.

\bibitem[\protect\citeauthoryear{Newton-Rex}{2012}]{jukedeck}
Ed~Newton-Rex.
\newblock Jukedeck.
\newblock \url{https://jukedeck.com/}, 2012.
\newblock Accessed: 2022-05-25.

\bibitem[\protect\citeauthoryear{Oore \bgroup \em et al.\egroup
  }{2020}]{oore2020time}
Sageev Oore, Ian Simon, Sander Dieleman, Douglas Eck, and Karen Simonyan.
\newblock This time with feeling: Learning expressive musical performance.
\newblock {\em Neural Computing and Applications}, 32(4):955--967, 2020.

\bibitem[\protect\citeauthoryear{Pachet}{2003}]{pachet2003continuator}
Francois Pachet.
\newblock The continuator: Musical interaction with style.
\newblock {\em Journal of New Music Research}, 32(3):333--341, 2003.

\bibitem[\protect\citeauthoryear{Pasquier \bgroup \em et al.\egroup
  }{2016}]{pasquier2016introduction}
Philippe Pasquier, Arne Eigenfeldt, Oliver Bown, and Shlomo Dubnov.
\newblock An introduction to musical metacreation.
\newblock {\em Computers in Entertainment (CIE)}, 14(2):2, 2016.

\bibitem[\protect\citeauthoryear{Payne}{2019}]{musenet}
Christine Payne.
\newblock Musenet.
\newblock \url{https://openai.com/blog/musenet/}, 2019.
\newblock Accessed: 2022-05-19.

\bibitem[\protect\citeauthoryear{Roberts \bgroup \em et al.\egroup
  }{2018}]{roberts2018hierarchical}
Adam Roberts, Jesse Engel, Colin Raffel, Curtis Hawthorne, and Douglas Eck.
\newblock A hierarchical latent vector model for learning long-term structure
  in music.
\newblock In {\em International conference on machine learning}, pages
  4364--4373. PMLR, 2018.

\bibitem[\protect\citeauthoryear{Roberts \bgroup \em et al.\egroup
  }{2019}]{roberts2019magenta}
Adam Roberts, Jesse Engel, Yotam Mann, Jon Gillick, Claire Kayacik, Signe
  N{\o}rly, Monica Dinculescu, Carey Radebaugh, Curtis Hawthorne, and Douglas
  Eck.
\newblock Magenta studio: Augmenting creativity with deep learning in ableton
  live.
\newblock In {\em 7th International Workshop on Musical Metacreation (MuMe
  2019)}, 2019.

\bibitem[\protect\citeauthoryear{Sturm \bgroup \em et al.\egroup
  }{2016}]{folkrnn}
Bob~L Sturm, Joao~Felipe Santos, Oded Ben-Tal, and Iryna Korshunova.
\newblock Music transcription modelling and composition using deep learning.
\newblock {\em arXiv preprint:1604.08723}, 2016.

\bibitem[\protect\citeauthoryear{Sturm \bgroup \em et al.\egroup
  }{2019}]{sturm2019machine}
Bob~L Sturm, Oded Ben-Tal, {\'U}na Monaghan, Nick Collins, Dorien Herremans,
  Elaine Chew, Ga{\"e}tan Hadjeres, Emmanuel Deruty, and Fran{\c{c}}ois Pachet.
\newblock Machine learning research that matters for music creation: A case
  study.
\newblock {\em Journal of New Music Research}, 48(1):36--55, 2019.

\bibitem[\protect\citeauthoryear{Sullivan}{2012}]{jnanalive}
Colin Sullivan.
\newblock Jnana live.
\newblock \url{https://ccrma.stanford.edu/~colinsul/projects/jnana/}, 2012.
\newblock Accessed: 2022-05-25.

\bibitem[\protect\citeauthoryear{Tchemeube \bgroup \em et al.\egroup
  }{2019}]{tchemeube2019apollo}
Renaud~Bougueng Tchemeube, Jeff Ens, and Philippe Pasquier.
\newblock Apollo: An interactive environment for generating symbolic musical
  phrases using corpus-based style imitation.
\newblock In {\em 7th International Workshop on Musical Metacreation (MuMe
  2019)}, 2019.

\bibitem[\protect\citeauthoryear{Tchemeube \bgroup \em et al.\egroup
  }{2022}]{bougueng2022calliope}
Renaud~Bougueng Tchemeube, Jeffrey~John Ens, and Philippe Pasquier.
\newblock Calliope: {A} co-creative interface for multi-track music generation.
\newblock In {\em C{\&}C '22: Creativity and Cognition, Venice, Italy, June 20
  - 23, 2022}, pages 608--611. {ACM}, 2022.

\bibitem[\protect\citeauthoryear{Thorogood and
  Pasquier}{2013}]{thorogood2013computationally}
Miles Thorogood and Philippe Pasquier.
\newblock Computationally created soundscapes with {A}udio {M}etaphor.
\newblock In {\em ICCC}, pages 1--7, 2013.

\bibitem[\protect\citeauthoryear{Velardo \bgroup \em et al.\egroup
  }{2014}]{melodrive}
Valerio Velardo, Andy Elmsley, Cárthach~Ó Nuanáin, and Christian Tronhjem.
\newblock Melodrive.
\newblock \url{https://melodrive.com/}, May 2014.
\newblock Accessed: 2022-05-25.

\bibitem[\protect\citeauthoryear{Yang}{2017}]{yang2017role}
Qian Yang.
\newblock The role of design in creating machine-learning-enhanced user
  experience.
\newblock In {\em 2017 AAAI Spring Symposium Series}, 2017.

\bibitem[\protect\citeauthoryear{Yang}{2018}]{yang2018machine}
Qian Yang.
\newblock Machine learning as a {U}{X} design material: How can we imagine
  beyond automation, recommenders, and reminders?
\newblock In {\em 2018 AAAI Spring Symposium Series}, 2018.

\end{thebibliography}

\end{document}